\documentclass[11pt,times]{article}

\usepackage{amsfonts}
\usepackage{amssymb,amsmath,amsthm}
\usepackage{epsfig}
\usepackage{color}
\usepackage{latexsym}
\usepackage{enumerate}
\usepackage{fullpage,setspace}
\usepackage[dvips, paper=letterpaper, top=1in, bottom=.75in, left=1in, right=1in, nohead, includefoot, footskip=.25in]{geometry}
\usepackage[backref,colorlinks,citecolor=blue,bookmarks=true]{hyperref}

\title{Sparse Covers for Sums of Indicators
}

\author{Constantinos Daskalakis\thanks{Supported by a Sloan Foundation fellowship, a Microsoft Research faculty fellowship and NSF Award CCF-0953960 (CAREER) and CCF-1101491.}\\EECS and CSAIL, MIT\\costis@mit.edu \and Christos Papadimitriou\thanks{Supported by NSF grant CCF-0964033 and a Google University Research Award.}\\Computer Science, U.C. Berkeley\\christos@cs.berkeley.edu}

\usepackage{hvdashln}
\usepackage{amsfonts}
\usepackage{amssymb,amsmath}
\usepackage{epsfig}
\usepackage{color}
\usepackage{latexsym}
\usepackage{enumerate}

\newtheorem{theorem}{Theorem}
\newtheorem{lemma}{Lemma}

\newtheorem{claim} {Claim}
\newtheorem{corollary}{Corollary}
\newtheorem{prop}{Proposition}

\newtheorem{remark}{Remark}

\newenvironment{prevproof}[2]{\noindent {\bf {Proof of {#1}~\ref{#2}:}}}{$\blacksquare$\vskip \belowdisplayskip}

\newcommand{\camera}[1]{}

\newcommand{\dtv}[2]{d_{\rm TV}\left(#1,#2\right)}

\def\H{\mathcal{H}}

\def\E{\mathbb{E}}
\def\Q{\mathcal{Q}}
\def\M{\mathcal{M}}

\def\R{\mathcal{R}}
\def\I{\mathcal{I}}

\def\L{\mathcal{L}}
\def\P{{\mathcal{P}}}
\def\B{{\mathcal{B}}}

\def\X{\mathcal{X}}
\def\Y{\mathcal{Y}}

\usepackage{color}
\definecolor{Red}{rgb}{1,0,0}

\begin{document}
\maketitle

\begin{abstract}
For all $n, \epsilon >0$, we show that the set of Poisson Binomial distributions on $n$ variables admits a  proper $\epsilon$-cover in total variation distance of size {$n^2+n \cdot (1/\epsilon)^{O(\log^2 (1/\epsilon))}$}, which can also be computed in polynomial time. We discuss the implications of our construction for approximation algorithms and the computation of approximate Nash equilibria in anonymous games. 
%
%
%

\end{abstract}

\section{Introduction}\label{sec:intro}

A {\em Poisson Binomial Distribution of order $n$} is the discrete probability distribution of the sum of $n$ independent indicator random variables. The distribution is parameterized by a vector $(p_i)_{i=1}^n \in [0, 1]^n$ of probabilities, and is denoted ${\rm PBD}(p_1, \ldots, p_n)$. In this paper we establish that the set ${\cal S}_n$ of all Poisson Binomial distributions of order $n$ admits certain useful covers with respect to the total variation distance $\dtv{\cdot}{\cdot}$ between distributions. Namely

\begin{theorem}[Main Theorem] \label{thm: sparse cover theorem}
For all $n, \epsilon >0$, there exists a set ${\cal S}_{n,\epsilon} \subset {\cal S}_n$ such that:

\begin{enumerate}
\item ${\cal S}_{n,\epsilon}$ is an $\epsilon$-cover of ${\cal S}_n$ in total variation distance; that is, for all $D \in {\cal S}_n$, there exists some $D' \in {\cal S}_{n,\epsilon}$ such that $\dtv{D}{D'} \le \epsilon$

\item {$|{\cal S}_{n,\epsilon}| \le n^2 + n \cdot \left({1 \over \epsilon}\right)^{O(\log^2{1/\epsilon})}$}

\item ${\cal S}_{n,\epsilon}$ can be computed in time {$O(n^2 \log n) + O(n \log n) \cdot \left({1 \over
\epsilon}\right)^{O(\log^2{1/\epsilon})}$.}
\end{enumerate}
Moreover, all distributions  ${\rm PBD}(p_1,\ldots,p_n)  \in {\cal S}_{n,\epsilon}$ in the cover satisfy at least one of the following properties, for some positive integer $k=k(\epsilon) = O(1/\epsilon):$ 
\begin{itemize}

\item ($k$-sparse form) there is some $ \ell \leq k^3$
such that, for all $i \leq \ell$, $p_i \in \left\{{1 \over k^2}, {2\over k^2},\ldots, {k^2-1 \over k^2 }\right\}$ and, for all $i >\ell $, $p_i \in \{0,  1\}$; or

\item ($(n,k)$-Binomial form) there is some {$\ell \in \{1,\dots,n\}$ and $q \in \left\{ {1 \over n}, {2 \over n},\ldots, {n \over n} \right\}$ such that, for all $i \leq \ell$, $p_i = q$ and, for all $i >\ell$, $p_i = 0$; moreover, $\ell$ and $q$ satisfy $\ell q \ge k^2$ and $\ell q(1-q) \ge k^2- k-1$.}

\end{itemize}
\end{theorem}

%
Covers such as the one provided by Theorem~\ref{thm: sparse cover theorem} are of interest in the design of algorithms, when one is searching a class of distributions $C$ to identify an element of the class with some quantitative property, or in optimizing over a class with respect to some objective. If the metric used in the construction of the cover is relevant for the problem at hand, and the cover is discrete, relatively small and easy to construct, then one can provide a useful approximation to the sought distribution by searching the cover, instead of searching all of $C$. For example, it is shown in~\cite{DaskalakisP07,DaskalakisP09,DaskalakisP13} that Theorem~\ref{thm: sparse cover theorem} implies efficient algorithms for computing approximate Nash equilibria in an important class of multiplayer games, called anonymous~\cite{Milchtaich96, Blonski99}.

\smallskip We proceed with a fairly detailed sketch of the proof of our main cover theorem, Theorem~\ref{thm: sparse cover theorem}, stating two additional results, Theorems~\ref{thm:larger cover} and~\ref{thm:moment matching}. The complete proofs of Theorems~\ref{thm: sparse cover theorem},~\ref{thm:larger cover} and~\ref{thm:moment matching} are deferred to Sections~\ref{sec:proof of sparse cover theorem},~\ref{sec: proof of weak cover theorem} and~\ref{sec: moment matching proof} respectively. Section~\ref{sec:related} discusses related work, while Section~\ref{sec:prelim} provides formal definitions, as well as known approximations to the Poisson Binomial distribution by simpler distributions, which are used in the proof.

\subsection{Proof Outline and Additional Results} \label{sec:outline of proof of sparse cover theorem}

At a high level, the proof of Theorem~\ref{thm: sparse cover theorem} is obtained in two steps. First, we establish the existence of an $\epsilon$-cover whose size is polynomial in $n$ and $(1/\epsilon)^{1/\epsilon^2}$, via Theorem~\ref{thm:larger cover}. We then show that this cover can be pruned to size polynomial in $n$ and $(1/\epsilon)^{\log^2(1/\epsilon)}$ using Theorem~\ref{thm:moment matching}, which provides a  quantification of how the total variation distance between Poisson Binomial distributions depends on the number of their first moments that are equal. 

We proceed to state the two ingredients of the proof, Theorems~\ref{thm:larger cover} and~\ref{thm:moment matching}. We start with Theorem~\ref{thm:larger cover} whose detailed sketch is given in Section~\ref{sec:outline of larger cover}, and complete proof in Section~\ref{sec: proof of weak cover theorem}. 
\begin{theorem} \label{thm:larger cover}
Let $X_1,\ldots,X_n$ be arbitrary mutually independent indicators, and $k \in \mathbb{N}$. Then there exist mutually independent indicators $Y_1,\ldots,Y_n$ satisfying the following:

\begin{enumerate}

\item $\dtv{\sum_{i}{X_i}}{\sum_{i}{Y_i}} \le 41/k ;$ \label{property_1}

\item at least one of the following is true:
\begin{enumerate}

\item  ($k$-sparse form) there exists some $ \ell \leq k^3$ such that, for all $i \leq \ell$, $\E[{Y_i}] \in \left\{{1 \over k^2}, {2\over k^2},\ldots, {k^2-1 \over k^2 }\right\}$ and, for all $i >\ell $, $\E[{Y_i}] \in \{0,  1\};$ or \label{enum:scenario2}

\item ($(n,k)$-Binomial form) there is some $\ell \in \{1,\dots,n\}$ and $q \in \left\{ {1 \over n}, {2 \over n},\ldots, {n \over n} \right\}$ such that, for all $i \leq \ell$, $\E[{Y_i}] = q$ and, for all $i >\ell$, $\E[{Y_i}] = 0;$ moreover, $\ell$ and $q$ satisfy $\ell q \ge k^2$ and $\ell q(1-q) \ge k^2- k-1.$ \label{enum:scenario1}

\end{enumerate}

\end{enumerate}
\end{theorem}

Theorem~\ref{thm:larger cover} implies the existence of an $\epsilon$-cover of ${\cal S}_n$  whose size is $n^2  + n \cdot \left({1 / \epsilon}\right)^{O({1/\epsilon^2})}$. This cover can be obtained by enumerating over all Poisson Binomial distributions of order $n$ that are in $k$-sparse or $(n,k)$-Binomial form as defined in the statement of the theorem, for $k=\lceil 41/ \epsilon \rceil$. 

 The next step is to sparsify this cover by removing elements to obtain Theorem~\ref{thm: sparse cover theorem}. Note that the term $n \cdot \left({1 / \epsilon}\right)^{O({1/\epsilon^2})}$ in the size of the cover is due to the enumeration over distributions in sparse form. Using Theorem~\ref{thm:moment matching} below, we argue that there is a lot of redundancy in those distributions, and that it suffices to only include $n\cdot \left({1 / \epsilon}\right)^{O(\log^2{1/\epsilon})}$ of them in the cover. In particular, Theorem~\ref{thm:moment matching} establishes that, if two Poisson Binomial distributions have their first $O(\log 1/\epsilon)$ moments equal, then their distance is at most $\epsilon$. So we only need to include at most one sparse form distribution with the same first $O(\log 1/\epsilon)$ moments in our cover. We proceed to state Theorem~\ref{thm:moment matching}, postponing its proof to Section~\ref{sec: moment matching proof}. In Section~\ref{sec:outline of moment matching} we provide a sketch of the proof.
 
\begin{theorem} \label{thm:moment matching}
Let $\mathcal{P}:=(p_i )_{i=1}^n \in [0,1/2]^n$ and
$\mathcal{Q}:=(q_i)_{i=1}^n \in [0,1/2]^n$ be two collections of
probability values. Let also
$\mathcal{X}:=(X_i)_{i=1}^n$ and $\mathcal{Y}:=(Y_i)_{i=1}^n$ be two
collections of mutually independent indicators with $\E[X_i]=p_i$ and
$\E[Y_i]=q_i$, for all $i \in [n]$. If for some $d \in [n]$ the
following condition is satisfied:
$$(C_d):~~\sum_{i=1}^n p_i^{\ell} = \sum_{i=1}^n q_i^{\ell},~~~\text{for all } \ell=1,\ldots,d,$$
\begin{align}
\text{then}~~\dtv{\sum_{i}{X_i}}{\sum_{i}{Y_i}}\le
13(d+1)^{1/4} 2^{-(d+1)/2}.~~\label{eq: target equation
X,Y}\end{align}
\end{theorem}
\begin{remark} \label{remark:equivalence of conditions} Condition $(C_d)$ in the statement of Theorem~\ref{thm:moment matching} constrains the first $d$ power sums of the expectations of the constituent indicators of two Poisson Binomial distributions. To relate these power sums to the moments of these distributions we can use the theory of symmetric polynomials to arrive at the following equivalent condition to $(C_d)$:
$$(V_d):~~\E\left[\left(\sum_{i=1}^n X_i\right)^{\ell}\right] = \E\left[\left(\sum_{i=1}^n Y_i\right)^{\ell}\right],~\text{for all } \ell \in [d].$$
We provide a proof that $(C_d) \Leftrightarrow (V_d)$ in Proposition~\ref{proposition:variable moments to probability moments} of Section~\ref{sec:missing proofs}.
\end{remark}
\begin{remark}
\smallskip In view of Remark~\ref{remark:equivalence of conditions}, Theorem~\ref{thm:moment matching} says the following: 

\medskip\begin{minipage}{15cm}{\em ``If two sums of independent indicators with expectations in [0,1/2] have equal first $d$ moments, then their total variation distance is $2^{-\Omega(d)}$.''} \end{minipage}

\medskip \noindent We note that the bound~\eqref{eq: target equation X,Y} does not depend on the number of variables $n$, and in particular does not rely on summing a large number of variables. We also note that, since we impose no constraint on the expectations of the indicators, we also impose no constraint on the variance of the resulting Poisson Binomial distributions. Hence we cannot use Berry-Ess\'een type bounds to bound the total variation distance of the two Poisson Binomial distributions by approximating them with Normal distributions. Finally, it is easy to see that Theorem~\ref{thm:moment matching} holds if we replace $[0,1/2]$ with $[1/2,1]$. See Corollary~\ref{theorem:binomial appx theorem for large guys} in Section~\ref{sec:missing proofs}.
%
\end{remark}

In Section~\ref{sec:proof of sparse cover theorem} we show how to use Theorems~\ref{thm:larger cover} and~\ref{thm:moment matching} to obtain Theorem~\ref{thm: sparse cover theorem}. We continue with the outlines of the proofs of Theorems~\ref{thm:larger cover} and~\ref{thm:moment matching}, postponing their complete proofs to Sections~\ref{sec: proof of weak cover theorem} and~\ref{sec: moment matching proof}.

\subsection{Outline of Proof of Theorem~\ref{thm:larger cover}} \label{sec:outline of larger cover}

Given arbitrary indicators $X_1,\ldots,X_n$ we obtain indicators $Y_1,\ldots,Y_n$, satisfying the requirements of Theorem~\ref{thm:larger cover}, in two steps. We first massage the given variables $X_1,\ldots,X_n$ to obtain variables $Z_1,\ldots,Z_n$ such that
\begin{align}
&\dtv{\sum_{i}{X_i}}{\sum_{i}{Z_i}} \le 7/k; \label{eq: target equation X,Z}\\
\text{and}~~~~&\E[Z_i] \notin \left(0,\frac{1}{k}\right)\cup\left(1-\frac{1}{k},1\right); \notag
\end{align}
that is, we eliminate from our collection variables that have expectations very close to $0$ or $1$, without traveling too much distance from the starting Poisson Binomial distribution.

Variables $Z_1,\ldots,Z_n$ do not necessarily satisfy Properties~\ref{enum:scenario2} or~\ref{enum:scenario1} in the statement of Theorem~\ref{thm:larger cover}, but allow us to define variables $Y_1,\ldots,Y_n$ which do satisfy one of these properties and, moreover,
\begin{align}
&\dtv{\sum_{i}{Z_i}}{\sum_{i}{Y_i}} \le 34/k. \label{eq: target equation Z,Y}
\end{align}
\eqref{eq: target equation X,Z}, \eqref{eq: target equation Z,Y} and the triangle inequality imply $\dtv{\sum_{i}{X_i}}{\sum_{i}{Y_i}} \le {41 \over k}$, concluding the proof of Theorem~\ref{thm:larger cover}.

\smallskip Let us call {\em Stage 1} the process of determining the $Z_i$'s and {\em Stage 2} the process of determining the $Y_i$'s. The two stages are described briefly below, and in detail in Sections~\ref{sec:stage1} and~\ref{sec:stage2} respectively. For convenience, we use the following notation: for $i=1,\ldots,n$, $p_i=\E[X_i]$ will denote the expectation of the given indicator $X_i$, $p_i' = \E[Z_i]$ the expectation of the intermediate indicator $Z_i$, and $q_i = \E[Y_i]$ the expectation of the final indicator $Y_i$.

\smallskip\noindent {\bf Stage 1:} Recall that our goal in this stage is to define a Poisson Binomial distribution $\sum_i Z_i$ whose constituent indicators have no expectation in ${\cal T}_k:=(0,\frac{1}{k})\cup(1-\frac{1}{k},1)$. The expectations $(p_i'=\E[Z_i])_i$ are defined in terms of the corresponding $(p_i)_i$ as follows. For all $i$, if $p_i \notin {\cal T}_k$ we set $p_i' = p_i$. Then, if ${\cal L}_k$ is the set of indices $i$ such that $p_i \in (0,1/k)$, we choose any collection $(p_i')_{i \in \L_k}$ so as to satisfy $|\sum_{i \in {\cal L}_k} p_i- \sum_{i \in {\cal L}_k} p_i'| \le 1/k$ and $p_i' \in \{0,1/k\}$, for all $i \in \L_k$. That is, we round all indicators' expectations to $0$ or $1/k$ while preserving the expectation of their sum, to within $1/k$. Using the Poisson approximation to the Poisson Binomial distribution, given as Theorem~\ref{lem:Poisson approximation} in Section~\ref{sec:approximations}, we can argue that $\sum_{i \in {\cal L}_k} X_i$ is within $1/k$ of a Poisson distribution with the same mean. By the same token, $\sum_{i \in {\cal L}_k} Z_i$ is $1/k$-close to a Poisson distribution with the same mean. And the two resulting Poisson distributions have means that are within $1/k$, and are therefore $1.5/k$-close to each other (see Lemma~\ref{lem:variation distance between Poisson distributions}). Hence, by triangle inequality $\sum_{i \in {\cal L}_k} X_i$ is $3.5/k$-close to $\sum_{i \in {\cal L}_k} Z_i$. A similar construction is used to define the $p_i'$'s corresponding to the $p_i$'s lying in $(1-1/k,1)$. The details of this step can be found in Section~\ref{sec:stage1}.

\smallskip \noindent {\bf Stage 2:} The definition of $(q_i)_i$ depends on the number $m$ of $p_i'$'s which are not $0$ or $1$. The case $m \le k^3$ corresponds to Case~\ref{enum:scenario2} in the statement of Theorem~\ref{thm:larger cover}, while the case $m > k^3$ corresponds to Case~\ref{enum:scenario1}. 

\begin{itemize}
\item Case $m \le k^3$:  First, we set $q_i = p_i'$, if $p_i' \in \{0,1\}$. We then argue that each $p'_i$, ${i \in \M}:=\{i~\vline~p_i' \notin \{0,1\} \}$, can be rounded to some $q_i$, which is an integer multiple of $1/k^2$, so that \eqref{eq: target equation Z,Y} holds. Notice that, if we were allowed to use multiples of $1/k^4$, this would be immediate via an application of Lemma~\ref{lem:coupling}:
$$\dtv{\sum_{i}{Z_i}}{\sum_{i}{Y_i}} \le \sum_{i \in \M} |p_i' - q_i|.$$
We improve the required accuracy to $1/k^2$ via a series of Binomial approximations to the Poisson Binomial distribution, using Ehm's bound~\cite{Ehm91} stated as Theorem~\ref{thm: binomial approximation to the poisson binomial distribution} in Section~\ref{sec:approximations}. The details involve partitioning the interval $[1/k,1-1/k]$ into irregularly sized subintervals, whose endpoints are integer multiples of $1/k^2$. We then round all but one of the $p_i'$'s falling in each subinterval to the endpoints of the subinterval so as to maintain their total expectation, and apply Ehm's approximation to argue that the distribution of their sum is not affected by more than $O(1/k^2)$ in total variation distance. It is crucial that the total number of subintervals is $O(k)$ to get a total hit of at most $O(1/k)$ in variation distance in the overall distribution. The details are given in Section~\ref{section: case m is small}.

\item Case $m > k^3$: We approximate $\sum_{i}Z_i$ with a Translated Poisson distribution (defined formally in Section~\ref{sec:prelim}), using Theorem~\ref{lem:translated Poisson approximation} of Section~\ref{sec:approximations} due to R\"ollin~\cite{Rollin07}. The quality of the approximation is inverse proportional to the standard deviation of $\sum_i Z_i$, which is at least $k$, by the assumption $m>k^3$. Hence, we show that $\sum_{i}Z_i$ is $3/k$-close to a Translated Poisson distribution. We then argue that the latter is $6/k$-close to a Binomial distribution $B(m',q)$, where {$m' \le n$ and $q$ is an integer multiple of $\frac{1}{n}$}. In particular, we show that an appropriate choice of $m'$ and $q$ implies~\eqref{eq: target equation Z,Y}, if we set $m'$ of the $q_i$'s  equal to $q$ and the remaining equal to $0$. The details are in Section~\ref{section: case m is large}.

\end{itemize}

\subsection{Outline of Proof of Theorem~\ref{thm:moment matching}} \label{sec:outline of moment matching}

Using Roos's expansion~\cite{Roos00}, given as Theorem~\ref{thm:roos} of Section~\ref{sec:approximations}, we express ${\rm PBD}(p_1,\ldots,p_n)$ as a weighted sum of the Binomial distribution $\B(n,p)$ at $p = \bar{p}= {\sum p_i / n}$ and its first $n$ derivatives with respect to $p$ also at value $p=\bar{p}$. (These derivatives correspond to finite signed measures.) We notice that the coefficients of the first $d+1$ terms of this expansion are symmetric polynomials in $p_1,\ldots,p_n$ of degree at most $d$. Hence, from the theory of symmetric polynomials, each of these coefficients can be written as a function of the power-sum symmetric polynomials $\sum_i p_i^{\ell}$ for $\ell=1,\ldots,d$. So, whenever two Poisson Binomial distributions satisfy Condition~$(C_d)$, the first $d+1$ terms of their expansions are exactly identical, and the total variation distance of the distributions depends only on the other terms of the expansion (those corresponding to higher derivatives of the Binomial distribution). The proof is concluded by showing that the joint contribution of these terms to the total variation distance can be bounded by $2^{-\Omega(d)}$, using Proposition~\ref{prop:roos truncation quality} of Section~\ref{sec:approximations}, which is also due to Roos~\cite{Roos00}. The details are provided in Section~\ref{sec: moment matching proof}.

\subsection{Related Work} \label{sec:related}
It is believed that Poisson~\cite{Poisson:37} was the first to study the Poisson Binomial distribution, hence its name. Sometimes the distribution is also referred to as ``Poisson's Binomial Distribution.'' PBDs have many uses in research areas such as survey sampling, case-control studies, and survival analysis; see~e.g.~\cite{ChenLiu:97} for a survey of their uses. They are also very important in the design of randomized algorithms~\cite{MR95}. 

In Probability and Statistics there is a broad literature studying various properties of these distributions; see \cite{Wang93} for an introduction to some of this work.  Many results provide approximations to the Poisson Binomial distribution via simpler distributions. In a well-known result, Le Cam~\cite{LeCam:60} shows that, for any vector $(p_i)_{i=1}^n \in [0, 1]^n$,
$$
d_{\rm TV} \left( {\rm PBD}(p_1, \ldots, p_n),{\rm Poisson} \left( \sum_{i=1}^n p_i \right) \right) \leq  \sum_{i=1}^n p_i^2,
$$
where ${\rm Poisson} (\lambda)$ is  the Poisson distribution with parameter $\lambda$.  Subsequently many other proofs of this bound and improved ones, such as Theorem~\ref{lem:Poisson approximation} of Section~\ref{sec:approximations}, were given, using a range of different techniques;
\cite{HodgesLC1960,Chen:74,barbour1984rate,DP:86} is a sampling of work along these lines, and Steele \cite{Steele:94} gives an extensive list of relevant references.    Much work has also been done on approximating PBDs by Normal distributions (see e.g. \cite{berry,esseen,Mikhailov:93,Volkova:95,chen2010normal}) and by Binomial distributions; see e.g. Ehm's result~\cite{Ehm91}, given as Theorem~\ref{thm: binomial approximation to the poisson binomial distribution} of Section~\ref{sec:approximations}, as well as~Soon's result~\cite{Soon:96} and Roos's result~\cite{Roos00}, given as Theorem~\ref{thm:roos} of Section~\ref{sec:approximations}. 

These results provide structural information about PBDs that can be well approximated by simpler distributions, but fall short of our goal of approximating a PBD to within {\em arbitrary accuracy}. Indeed, the approximations obtained in the probability literature (such as the Poisson, Normal and Binomial approximations)  typically depend on the first few moments of the  PBD being approximated, while higher moments are crucial for arbitrary approximation~\cite{Roos00}. At the same time, algorithmic applications often require that the approximating distribution is of the same kind as the distribution that is being approximated. E.g., in the anonymous game application mentioned earlier, the parameters of the given PBD correspond to mixed strategies of players at Nash equilibrium, and the parameters of the approximating PBD correspond to mixed strategies at approximate Nash equilibrium. Approximating the given PBD via a Poisson or a Normal distribution would not have any meaning in the context of a game. 

{As outlined above, the proof of our main result, Theorem~\ref{thm: sparse cover theorem}, builds on Theorems~\ref{thm:larger cover} and~\ref{thm:moment matching}. A weaker form of these theorems was announced in~\cite{Daskalakis08,DaskalakisP09}, while a weaker form of Theorem~\ref{thm: sparse cover theorem} was announced in~\cite{DaskalakisDS12}.}


\section{Preliminaries} \label{sec:prelim}

For a positive integer $\ell$, we denote by $[\ell]$ the set $\{1,\dots,\ell\}$. For a random variable $X$, we denote by ${\cal L}(X)$ its distribution. We further need the following definitions.

\medskip \noindent {\em Total variation distance:} For two distributions $\mathbb{P}$ and $\mathbb{Q}$ supported on a finite set ${A}$ their {\em total variation distance} is defined as
$$\dtv{\mathbb{P}}{\mathbb{Q}} := \frac{1}{2} \sum_{\alpha \in A}{\left|\mathbb{P}(\alpha)-\mathbb{Q}(\alpha)\right|}.$$
An equivalent way to define $\dtv{\mathbb{P}}{\mathbb{Q}}$ is to view $\mathbb{P}$ and $\mathbb{Q}$ as vectors in $\mathbb{R}^{A}$, and define $\dtv{\mathbb{P}}{\mathbb{Q}} = {1 \over 2}\|\mathbb{P} - \mathbb{Q}\|_1$ to equal half of their $\ell_1$ distance. 
If $X$ and $Y$ are random variables ranging over a finite set, their total variation distance, denoted $\dtv{X}{Y},$ is defined to equal $\dtv{{\cal L}(X)}{{\cal L}(Y)}$. 

\medskip \noindent {\em Covers:} Let ${\cal F}$ be a set of probability distributions. A subset ${\cal G} \subseteq {\cal F}$ is called a {\em (proper) $\epsilon$-cover} of ${\cal F}$ in total variation distance if, for all $D \in {\cal F}$, there exists some $D' \in {\cal G}$ such that $\dtv{D}{D'} \le \epsilon$.

\medskip \noindent {\em Poisson Binomial Distribution:} A {\em Poisson Binomial distribution of order $n \in \mathbb{N}$} is the discrete probability distribution of the sum $\sum_{i=1}^n X_i$ of $n$ mutually independent Bernoulli random variables $X_1,\ldots,X_n$. We denote the set of all Poisson Binomial distributions of order $n$ by ${\cal S}_{n}$. 

By definition, a Poisson Binomial distribution $D  \in {\cal S}_n$ can be represented by a vector  $(p_i)_{i=1}^n \in [0,1]^n$ of probabilities as follows. We map $D \in {\cal S}_n$ to  a vector of probabilities by finding a collection $X_1,\ldots,X_n$ of mutually independent indicators such that $\sum_{i=1}^n X_i$ is distributed according to $D$, and setting $p_i = \E[X_i]$ for all~$i$. The following lemma implies that the resulting vector of probabilities is unique up to a permutation, so that there is a one-to-one correspondence between Poisson Binomial distributions and vectors $(p_i)_{i=1}^n \in [0,1]^n$ such that $0\le p_1 \le p_2 \le \ldots \le p_n \le 1$. The proof of this lemma can be found in Section~\ref{sec:missing proofs}.
 
\begin{lemma} \label{lem: PBDs are fully determined by their probability vectors}
Let $X_1,\ldots,X_n$ be mutually independent indicators with expectations $p_1 \le p_2 \le \ldots \le p_n$ respectively. Similarly let $Y_1,\ldots,Y_n$ be mutually independent indicators with expectations $q_1 \le  \ldots \le q_n$ respectively. The distributions of $\sum_i X_i$ and $\sum_i Y_i$ are different if and only if $(p_1,\ldots,p_n) \neq (q_1,\ldots,q_n)$.
\end{lemma}

\smallskip We will be denoting a Poisson Binomial distribution  $D  \in {\cal S}_n$ by ${\rm PBD}(p_1,\ldots,p_n)$ when it is the distribution of the sum $\sum_{i=1}^n X_i$ of mutually independent indicators  $X_1,\ldots,X_n$ with expectations $p_i=\E[X_i]$, for all $i$. Given the above discussion, the representation is unique up to a permutation of the $p_i$'s.

\medskip \noindent {\em Translated Poisson Distribution:} We say that an integer random variable $Y$ has a {\em translated Poisson distribution} with parameters $\mu$ and $\sigma^2$ and write $\L(Y)=TP(\mu,\sigma^2)$
iff $$\L(Y - \lfloor \mu-\sigma^2\rfloor) = {\rm Poisson}(\sigma^2+ \{\mu-\sigma^2\}),$$ where $\{\mu-\sigma^2\}$ represents the fractional part of $\mu-\sigma^2$.

\medskip \noindent {\em {Order Notation}:} Let $f(x)$ and $g(x)$ be two positive functions defined on some infinite subset of $\mathbb{R}_+$. One writes $f(x)=O(g(x))$ if and only if, for sufficiently large values of $x$, $f(x)$ is at most a constant times $g(x)$. That is, $f(x) = O(g(x))$ if and only if there exist positive real numbers $M$ and $x_0$ such that
$$    f(x) \le \; M g(x),\mbox{ for all }x>x_0.$$
Similarly, we write $f(x) = \Omega(g(x))$ if and only if there exist positive reals $M$ and $x_0$ such that
$$    f(x) \ge \; M g(x),\mbox{ for all }x>x_0.$$

We are casual in our use of the order notation $O(\cdot)$ and $\Omega(\cdot)$ throughout the paper. Whenever we write $O(f(n))$ or $\Omega(f(n))$ in some bound where $n$ ranges over the integers, we mean that there exists a constant $c >0 $ such that the bound holds true for sufficiently large $n$ if we replace the $O(f(n))$ or $\Omega(f(n))$ in the bound by $c \cdot f(n)$. On the other hand, whenever we write $O(f(1/\epsilon))$ or $\Omega(f(1/\epsilon))$ in some bound where $\epsilon$ ranges over the positive reals, we mean that there exists a constant $c>0$ such that the bound holds true for sufficiently {\em small} $\epsilon$ if we replace the $O(f(1/\epsilon))$ or $\Omega(f(1/\epsilon))$ in the bound with $c \cdot f(1/\epsilon)$.


\medskip We conclude with an easy but useful lemma whose proof we defer to Section~\ref{sec:missing proofs}.
\begin{lemma} \label{lem:coupling}
Let $X_1,\ldots,X_n$ be mutually independent random variables, and let $Y_1,\ldots,Y_n$ be mutually independent random variables. Then
$$\dtv{\sum_{i=1}^n X_i}{\sum_{i=1}^n Y_i} \le \sum_{i=1}^n\dtv{X_i}{Y_i}.$$
\end{lemma}

\subsection{Approximations to the Poisson Binomial Distribution} \label{sec:approximations}

We present a collection of known approximations to the Poisson Binomial distribution via simpler distributions. The quality of these approximations can be quantified in terms of the first few moments of the Poisson Binomial distribution that is being approximated. We will make use of these bounds to approximate Poisson Binomial distributions in different regimes of their moments. Theorems~\ref{lem:Poisson approximation}---\ref{lem:translated Poisson approximation} are obtained via the Stein-Chen method.

\begin{theorem}[Poisson Approximation~\cite{barbour1984rate,BarbourEtAl:book}] \label{lem:Poisson approximation}
Let $J_1,\ldots,J_n$ be mutually independent indicators with $\E[J_i]=t_i$. Then
$$\dtv{\sum_{i=1}^nJ_i}{{\rm Poisson}\left(\sum_{i=1}^nt_i\right)} \le \frac{\sum_{i=1}^nt_i^2}{\sum_{i=1}^nt_i}.$$
\end{theorem}

\begin{theorem}[Binomial Approximation~\cite{Ehm91}]\label{thm: binomial approximation to the poisson binomial distribution}
Let $J_1,\ldots,J_n$ be mutually independent indicators with $\E[J_i]=t_i$, and $\bar{t} = {\sum_i t_i \over n}$. Then
$$\dtv{\sum_{i=1}^nJ_i}{\B\left(n,\bar{t}\right)} \le \frac{\sum_{i=1}^n (t_i-\bar{t})^2}{(n+1) \bar{t}(1-\bar{t})},$$
where $\B\left(n,\bar{t}\right)$ is the Binomial distribution with parameters $n$ and $\bar{t}$. 
\end{theorem}

\begin{theorem}[Translated Poisson Approximation\cite{Rollin07}] \label{lem:translated Poisson approximation}
Let $J_1,\ldots,J_n$ be mutually independent indicators with $\E[J_i]=t_i$. Then
$$\dtv{\sum_{i=1}^nJ_i}{TP(\mu, \sigma^2)} \le \frac{\sqrt{\sum_{i=1}^nt_i^3(1-t_i)}+2}{\sum_{i=1}^nt_i(1-t_i)},$$
where $\mu=\sum_{i=1}^nt_i$ and $\sigma^2 = \sum_{i=1}^nt_i(1-t_i)$.
\end{theorem}

The approximation theorems stated above do not always provide tight enough approximations. When these fail, we employ the following theorem of Roos~\cite{Roos00}, which provides an expansion of the Poisson Binomial distribution as a weighted sum of a finite number of signed measures: the Binomial distribution $\mathcal{B}({n,p})$ (for an arbitrary value of $p$) and its first  $n$ derivatives with respect to the parameter $p$, at the chosen value of $p$. For the purposes of the following statement we denote by $\mathcal{B}_{n,p}(m)$ the probability assigned by the Binomial distribution $\B(n,p)$ to integer $m$.

\begin{theorem}[\cite{Roos00}] \label{thm:roos}
Let $\mathcal{P}:=(p_i )_{i=1}^n \in [0,1]^n$, $X_1,\ldots,X_n$ be mutually
independent indicators with expectations $p_1,\ldots,p_n$, and  $X=\sum_i X_i$. Then, for all $m \in \{0,\ldots,n\}$ and  $p
\in [0,1]$,
\begin{align}
Pr[X = m]  = \sum_{\ell = 0}^n \alpha_{\ell}(\mathcal{P}, p)\cdot
\delta^{\ell}\mathcal{B}_{n,p}(m), \label{eq:krawtchouk expansion}
\end{align}
where for the purposes of the above expression:
\begin{itemize}
\item $\alpha_0(\mathcal{P},p):=1$ and for $\ell \in [n]:$
$$\alpha_{\ell}(\mathcal{P},p):= \sum_{1 \le k(1) < \ldots < k(\ell)
\le n} \prod_{r=1}^{\ell}(p_{k(r)}-p);$$
\item and for all $\ell \in \{0,\ldots,n\}:$
$$\delta^{\ell}\mathcal{B}_{n,p}(m):=\frac{(n-\ell)!}{n!} \frac{d^{\ell}}{d p^{\ell}}\mathcal{B}_{n,p}(m),$$
where for the last definition we interpret $\mathcal{B}_{n,p}(m)\equiv{n \choose m} p^m (1-p)^{n-m}$ as a function of $p$.
\end{itemize}
\end{theorem}
\noindent We can use Theorem~\ref{thm:roos} to get tighter approximations to the Poisson Binomial distribution by appropriately tuning the number of terms of summation~\eqref{eq:krawtchouk expansion} that we keep. The following proposition, shown in the proof of Theorem 2 of~\cite{Roos00}, bounds the $\ell_1$ approximation error to the Poisson Binomial distribution when only the first $d+1$ terms of summation~\eqref{eq:krawtchouk expansion} are kept. The error decays exponentially in $d$ as long as the quantity  $\theta(\P,p)$ in the proposition statement is smaller than $1$.
\begin{prop}[\cite{Roos00}] \label{prop:roos truncation quality}
Let $\mathcal{P}=(p_i )_{i=1}^n \in [0,1]^n$, $p \in [0,1]$, $\alpha_{\ell}(\cdot, \cdot)$ and $\delta^{\ell}\mathcal{B}_{n,p}(\cdot)$ as in the statement of Theorem~\ref{thm:roos}, and take
$$\theta(\P,p)= \frac{2 \sum_{i=1}^n(p_i - p)^2 + (\sum_{i=1}^n(p_i - p))^2}{2np(1-p)}.$$
If  $\theta(\P,p)< 1,$ then, for all $d \ge 0$:
\begin{align*}
&\sum_{\ell = d+1}^n |\alpha_{\ell}(\mathcal{P}, p)|\cdot
\| \delta^{\ell}\mathcal{B}_{n,p}(\cdot)\|_1 \le
{\sqrt{e}(d+1)^{1/4}} \theta(\P,p)^{(d+1)/2} \frac{1-
\frac{d}{d+1}\sqrt{\theta(\P,p)}}{(1-\sqrt{\theta(\P,p)})^2},
\end{align*}
where $\| \delta^{\ell}\mathcal{B}_{n,p}(\cdot)\|_1 := \sum_{m=0}^n | \delta^{\ell}\mathcal{B}_{n,p}(m) |$.
\end{prop}

\section{Proof of Theorem~\ref{thm: sparse cover theorem}} \label{sec:proof of sparse cover theorem}

We first argue that Theorem~\ref{thm:larger cover} already implies the existence of an $\epsilon$-cover ${\cal S}_{n,\epsilon}'$ of ${\cal S}_n$ of size at most $n^2  + n \cdot \left({1 \over \epsilon}\right)^{O({1/\epsilon^2})}$. This cover is obtained by taking the union of all Poisson Binomial distributions in $(n,k)$-Binomial form and all Poisson Binomial distributions in $k$-sparse form, for $k=\lceil 41/ \epsilon \rceil$. The total number of Poisson Binomial distributions in $(n,k)$-Binomial form is at most $n^2$, since there are at most $n$ choices for the value of $\ell$ and at most $n$ choices for the value of $q$. The total number of Poisson Binomial distributions  in $k$-sparse form is at most $(k^3+1)\cdot k^{3 k^2} \cdot (n+1) =  n \cdot \left({1 \over \epsilon}\right)^{O({1/\epsilon^2})}$ since there are $k^3+1$ choices for $\ell$, at most $k^{3k^2}$ choices of probabilities $p_1 \le p_2 \le \ldots \le p_{\ell}$ in $\left\{{1 \over k^2}, {2\over k^2},\ldots, {k^2-1 \over k^2 }\right\}$, and at most $n+1$ choices for the number of variables indexed by $i > \ell$ that have expectation equal to $1$.\footnote{Note that imposing the condition $p_1 \le \ldots \le p_{\ell}$ won't lose us any Poisson Binomial distribution in $k$-sparse form given Lemma~\ref{lem: PBDs are fully determined by their probability vectors}.} Notice that enumerating over the above distributions takes time $O(n^2 \log n)  + O(n \log n) \cdot \left({1 \over \epsilon}\right)^{O({1/\epsilon^2})}$, as a number in $\{0,\ldots,n\}$ and a probability in $\left\{{1 \over n},{2 \over n},\ldots,{n \over n}\right\}$ can be represented using $O(\log n)$ bits, while a number in $\{0,\ldots,k^3\}$ and a probability in $\left\{{1 \over k^2}, {2\over k^2},\ldots, {k^2-1 \over k^2 }\right\}$ can be represented using $O(\log k)=O(\log 1/\epsilon)$ bits.

\smallskip We next show that we can remove from ${\cal S}_{n,\epsilon}'$ a large number of the sparse-form distributions  it contains to obtain a $2\epsilon$-cover of ${\cal S}_n$. In particular, we shall only keep $n \cdot \left({1 \over \epsilon}\right)^{O(\log^2{1/\epsilon})}$ sparse-form distributions by appealing to Theorem~\ref{thm:moment matching}. To explain the pruning we introduce some notation. For a collection $\P=(p_i)_{i \in [n]} \in [0,1]^n$ of probability values we denote by ${\cal L}_{\P} =\{i~|~p_i\in (0,1/2]\}$ and by ${\cal R}_{\P}= \{i~|~p_i\in (1/2,1)\}$. Theorem~\ref{thm:moment matching}, Corollary~\ref{theorem:binomial appx theorem for large guys}, Lemma~\ref{lem:coupling} and Lemma~\ref{lem: PBDs are fully determined by their probability vectors} imply  that if two collections $\P=(p_i)_{i \in [n]}$ and $\Q=(q_i)_{i \in [n]}$ of probability values satisfy
\begin{align*}
\sum_{i \in {\cal L}_{\P}} p_i^{t} &= \sum_{i \in {\cal L}_{\Q}} q_i^{t},~~~\text{for all } t=1,\ldots,d; \\
\sum_{i \in {\cal R}_{\P}} p_i^{t} &= \sum_{i \in {\cal R}_{\Q}} q_i^{t},~~~\text{for all } t=1,\ldots,d; \text{and}\\
(p_i)_{[n]\setminus (\L_{\P} \cup \R_{\P})}~\text{and}&~(q_i)_{[n]\setminus (\L_{\Q} \cup \R_{\Q})}~\text{are equal up to a permutation;}
\end{align*}
then $d_{\rm TV}({\rm PBD}(\P), {\rm PBD}(\Q)) \le 2\cdot 13(d+1)^{1/4} 2^{-(d+1)/2}$. In particular, for some $d(\epsilon)= O(\log 1/\epsilon)$, this bound becomes at most $\epsilon$. 

For a collection $\P=(p_i)_{i \in [n]} \in [0,1]^n$, we define its {\em moment profile $m_{\P}$} to be the $(2 d(\epsilon)+1)$-dimensional vector 
$$m_{\P} = \left(\sum_{i \in {\cal L}_{\P}} p_i, \sum_{i \in {\cal L}_{\P}} p_i^{2},\ldots,\sum_{i \in {\cal L}_{\P}} p_i^{d(\epsilon)}; \sum_{i \in {\cal R}_{\P}} p_i, \ldots,\sum_{i \in {\cal R}_{\P}} p_i^{d(\epsilon)} ; |\{ i~|~p_i=1 \}|\right).$$
By the previous discussion, for two collections $\P, \Q$, if $m_{\P} = m_{\Q}$ then $d_{\rm TV}({\rm PBD}(\P), {\rm PBD}(\Q)) \le \epsilon$.

Given the above we sparsify ${\cal S}_{n,\epsilon}'$ as follows: for every possible moment profile that can arise from a Poisson Binomial distribution in $k$-sparse form, we keep in our cover a single Poisson Binomial distribution with such moment profile. The cover resulting from this sparsification is a $2 \epsilon$-cover, since the sparsification loses us an additional $\epsilon$ in total variation distance, as argued above. 

We now bound the cardinality of the sparsified cover. The total number of moment profiles of  $k$-sparse Poisson Binomial distributions is 
$k^{O(d(\epsilon)^2)} \cdot (n+1)$.
Indeed, consider a Poisson Binomial distribution ${\rm PBD}(\P=(p_i)_{i \in [n]})$ in $k$-sparse form. There are at most  $k^3+1$ choices for $|{\cal L}_{\P}|$, at most $k^3+1$ choices for $|{\cal R}_{\P}|$, and at most $(n+1)$ choices for $|\{i~|~p_i = 1\}|$. We also claim that the total number of possible vectors 
$$\left(\sum_{i \in {\cal L}_{\P}} p_i, \sum_{i \in {\cal L}_{\P}} p_i^{2},\ldots,\sum_{i \in {\cal L}_{\P}} p_i^{d(\epsilon)}\right)$$
is $k^{O(d(\epsilon)^2)}$. Indeed, if $|\L_{\P}|=0$ there is just one such vector, namely the all-zero vector. If $|\L_{\P}|> 0$, then, for all $t=1,\ldots,d(\epsilon)$, $\sum_{i \in {\cal L}_{\P}} p_i^{t} \in (0, |{\cal L}_{\P}|]$ and it must be an integer multiple of $1/k^{2t}$. So the total number of possible values of $\sum_{i \in {\cal L}_{\P}} p_i^{t}$ is at most $k^{2t} |{\cal L}_{\P}| \le k^{2t} k^3$, and the total number of possible vectors $$\left(\sum_{i \in {\cal L}_{\P}} p_i, \sum_{i \in {\cal L}_{\P}} p_i^{2},\ldots,\sum_{i \in {\cal L}_{\P}} p_i^{d(\epsilon)}\right)$$ is at most
$$\prod_{t=1}^{d(\epsilon)}k^{2t} k^3 \le k^{O(d(\epsilon)^2)}.$$
The same upper bound applies to the total number of possible vectors
$$\left(\sum_{i \in {\cal R}_{\P}} p_i, \sum_{i \in {\cal R}_{\P}} p_i^{2},\ldots,\sum_{i \in {\cal R}_{\P}} p_i^{d(\epsilon)}\right).$$
The moment profiles we enumerated over are a superset of the moment profiles of $k$-sparse Poisson Binomial distributions. We call them {\em compatible} moment profiles. We argued that there are at most $k^{O(d(\epsilon)^2)} \cdot (n+1)$ compatible moment profiles, so the total number of Poisson Binomial distributions in $k$-sparse form that we keep in the cover is at most $k^{O(d(\epsilon)^2)} \cdot (n+1) =  n \cdot \left({1 \over \epsilon}\right)^{O(\log^2{1/\epsilon})}$. The number of Poisson Binomial distributions in $(n,k)$-Binomial form is the same as before, i.e. at most $n^2$, as we did not eliminate any of them. So the size of the sparsified cover is $n^2 + n \cdot \left({1 \over \epsilon}\right)^{O(\log^2{1/\epsilon})}$.

\smallskip To finish the proof it remains to argue that we don't actually need to first compute ${\cal S}_{n,\epsilon}'$ and then sparsify it to obtain our cover, but can produce it directly in time $O(n^2 \log n) + O(n \log n) \cdot \left({1 \over \epsilon}\right)^{O(\log^2{1/\epsilon})}$. We claim that, given a moment profile $m$ that is compatible with a $k$-sparse Poisson Binomial distribution, we can compute some ${\rm PBD}(\P=(p_i)_i)$ in $k$-sparse form such that $m_{\P}=m$, if such a distribution exists, in time $O(\log n) \left({1 \over \epsilon}\right)^{O(\log^2{1/\epsilon})}$. This follows from Claim~\ref{claim: easy to solve moment equations} of Section~\ref{sec:missing proofs}.\footnote{A naive application of Claim~\ref{claim: easy to solve moment equations} results in running time $O(n^3 \log n) \cdot \left({1 \over \epsilon}\right)^{O(\log^2{1/\epsilon})}$. We can improve this to the claimed running time as follows: for all possible values $|{\cal L}_{\P}|$, $|{\cal R}_{\P}|$ such that $|{\cal L}_{\P}| + |{\cal R}_{\P}| \le \min(k^3,n-m_{2d(\epsilon)+1})$, we invoke Claim~\ref{claim: easy to solve moment equations} with $\tilde{n}=|{\cal L}_{\P}| + |{\cal R}_{\P}|$, $\delta=d(\epsilon)$, $B=k^3$, $n_0=n_1=0$, $n_s=|{\cal L}_{\P}|$, $n_b=|{\cal R}_{\P}|$, and moments $\mu_\ell = m_{\ell}$, for $\ell=1,\ldots,d(\epsilon)$, and $\mu_\ell' = m_{d(\epsilon)+\ell}$, for $\ell=1,\ldots,d(\epsilon)$. If for some pair $|{\cal L}_{\P}|$, $|{\cal R}_{\P}|$ the algorithm succeeds in finding probabilities matching the provided moments, we set $m_{2d(\epsilon)+1}$ of the remaining probabilities equal to $1$ and the rest  to $0$. Otherwise, we output ``fail.''} So our algorithm enumerates over all moment profiles that are compatible with a $k$-sparse Poisson Binomial distribution and for each profile invokes Claim~\ref{claim: easy to solve moment equations} to find a Poisson Binomial distribution with such moment profile, if such distribution exists, adding it to the cover if it does exist. It then enumerates over all Poisson Binomial distributions in $(n,k)$-Binomial form and adds them to the cover as well. The overall running time is as promised.

\section{Proof of Theorem~\ref{thm:larger cover}} \label{sec: proof of weak cover theorem}

We organize the proof according to the structure and notation of our outline in Section~\ref{sec:outline of larger cover}. In particular, we proceed to provide the details of Stages 1 and 2, described in the outline. The reader should refer to Section~\ref{sec:outline of larger cover} for notation.
\subsection{Details of Stage 1} \label{sec:stage1}
Define
${\cal L}_k:= \left\{i~\vline~i\in[n] \wedge p_i\in(0,1/k)\right\}$ and ${\cal H}_k:= \left\{i~\vline~i\in[n] \wedge p_i\in(1-1/k,1)\right\}.$ We define the expectations $(p_i')_i$ of the intermediate indicators $(Z_i)_i$ as follows. 

\medskip First, we set $p'_i=p_i$, for all $i \in [n]\setminus{\cal L}_k \cup {\cal H}_k$. It follows that
\begin{align}
&\dtv{\sum_{i \in [n]\setminus{\cal L}_k \cup {\cal H}_k}X_i}{\sum_{i \in [n]\setminus{\cal L}_k \cup {\cal H}_k}Z_i} = 0.\label{eq:weeding out approximation intermediate}
\end{align}\\

\smallskip Next, we define the probabilities $p'_i$, $i\in{\cal L}_k$, using the following procedure:
\begin{enumerate}
\item Set $r=\left\lfloor \frac{\sum_{i \in {\cal L}_k}p_i}{1/k}\right\rfloor$; and let ${\cal L}_k' \subseteq {\cal L}_k$ be an arbitrary subset of cardinality $|{\cal L}_k'|=r$.
\item Set $p_i'=\frac{1}{k}$, for all $i \in {\cal L}_k'$, and $p_i'=0$, for all $i \in {\cal L}_k\setminus {\cal L}_k'$.
\end{enumerate}
We bound the total variation distance $\dtv{\sum_{i \in {\cal L}_k}X_i}{\sum_{i \in {\cal L}_k}Z_i}$ using the Poisson approximation to the Poisson Binomial distribution. In particular, Theorem~\ref{lem:Poisson approximation} implies
$$\dtv{\sum_{i \in {\cal L}_k}X_i}{{\rm Poisson}\left(\sum_{i \in \L_k} p_i\right)} \le \frac{\sum_{i \in \L_k}p_i^2}{\sum_{i \in \L_k} p_i} \le {{1 \over k} \sum_{i \in \L_k} p_i \over \sum_{i \in \L_k} p_i}= 1/k.$$
Similarly, $\dtv{\sum_{i \in {\cal L}_k}Z_i}{{\rm Poisson}\left(\sum_{i \in \L_k} p'_i\right)} \le 1/k.$ Finally, we use Lemma~\ref{lem:variation distance between Poisson distributions} (given below and proved in Section~\ref{sec:missing proofs}) to bound the distance
$$\dtv{{\rm Poisson}\left(\sum_{i \in \L_k} p_i\right)}{{\rm Poisson}\left(\sum_{i \in \L_k} p'_i\right)} \le {1\over 2}\left(e^{1\over k}-e^{-{1 \over k}} \right)\le {1.5 \over k},$$
where we used that $|\sum_{i \in \L_k} p_i - \sum_{i \in \L_k} p'_i | \le 1/k$. Using the triangle inequality the above imply
\begin{align}
&\dtv{\sum_{i \in {\cal L}_k}X_i}{\sum_{i \in {\cal L}_k}Z_i} \le \frac{3.5}{k}.\label{eq:weeding out approximation low}
\end{align}

\begin{lemma}[Variation Distance of Poisson Distributions] \label{lem:variation distance between Poisson distributions}
Let  $\lambda_1, \lambda_2>0$ . Then
\begin{align*}&\dtv{{\rm Poisson}(\lambda_1)}{{\rm Poisson}(\lambda_2)} \le {1 \over 2} \left(e^{|\lambda_1-\lambda_2|}-e^{-|\lambda_1-\lambda_2|}\right).\end{align*}
\end{lemma}

\smallskip  We follow a similar rounding scheme to define $(p_i')_{i \in \H_k}$ from $(p_i)_{i \in {\cal H}_k}$. That is, we round some of the $p_i$'s to $1-1/k$ and some of them to $1$ so that $|\sum_{i \in \H_k}p_i -\sum_{i \in \H_k}p'_i | \le 1/k$. As a result, we get (to see this, repeat the argument employed above to the variables $1-X_i$ and $1-Z_i$, $i\in {\cal H}_k$)
\begin{align}
&\dtv{\sum_{i \in {\cal H}_k}X_i}{\sum_{i \in {\cal H}_k}Z_i} \le \frac{3.5}{k}.\label{eq:weeding out approximation high}
\end{align}
Using \eqref{eq:weeding out approximation intermediate},~\eqref{eq:weeding out approximation low},~\eqref{eq:weeding out approximation high} and Lemma~\ref{lem:coupling} we get~\eqref{eq: target equation X,Z}.

\subsection{Details of Stage 2}\label{sec:stage2}
Recall that ${\cal M}:=\{i~|~p'_i \notin \{0,1\}\}$ and $m:=|{\cal M}|$. Depending on on whether $m \le k^3$ or $m > k^3$ we follow different strategies to define  the expectations $(q_i)_i$ of indicators $(Y_i)_i$.

\subsubsection{The Case $m \le k^3$} \label{section: case m is small}

\medskip First we set $q_i = p'_i$, for all $i\in [n] \setminus {\cal M}.$ It follows that
\begin{align}
&\dtv{\sum_{i \in [n] \setminus {\cal M}}Z_i}{\sum_{i \in [n] \setminus {\cal M}}Y_i} = 0.\label{eq:stage 2 akrianoi}
\end{align}

\medskip For the definition of $(q_i)_{i \in \M}$, we make use of Ehm's Binomial approximation to the Poisson Binomial distribution, stated as Theorem~\ref{thm: binomial approximation to the poisson binomial distribution} in Section~\ref{sec:approximations}. We start by partitioning $\M$ as $\M=\M_l \sqcup \M_h$, where $\M_l = \{i \in \M~|~p'_i \le 1/2 \}$, and describe below a procedure for defining $(q_i)_{i \in \M_l}$ so that the following hold:
\begin{enumerate}
\item $\dtv{ \sum_{i \in \M_l} Z_i}{\sum_{i \in \M_l}Y_i} \le 17/k$; \label{item: property 1}
\item for all $i \in \M_l$, $q_i$ is an integer multiple of $1/k^2$. \label{item:property 2}
\end{enumerate}
To define $(q_i)_{i \in \M_h}$, we apply the same procedure to $(1-p_i')_{i\in \M_h}$ to obtain $(1-q_i)_{i \in \M_h}$. Assuming the correctness of our procedure for probabilities $\le 1/2$ the following should also hold:
\begin{enumerate}
\item $\dtv{\sum_{i \in \M_h} Z_i}{\sum_{i \in \M_h}Y_i} \le 17/k$;
\item for all $i \in \M_h$, $q_i$ is an integer multiple of $1/k^2$.
\end{enumerate}
Using Lemma~\ref{lem:coupling}, the above bounds imply 
\begin{align}
\dtv{ \sum_{i \in \M} Z_i}{\sum_{i \in \M}Y_i} \le \dtv{\sum_{i \in \M_l} Z_i}{\sum_{i \in \M_l}Y_i} + \dtv{\sum_{i \in \M_h} Z_i}{\sum_{i \in \M_h}Y_i}\le 34/k.\label{eq: small m final}
\end{align}
Now that we have~\eqref{eq: small m final}, using~\eqref{eq:stage 2 akrianoi} and Lemma~\ref{lem:coupling} we get~\eqref{eq: target equation Z,Y}.

\medskip So it suffices to define the $(q_i)_{i \in \M_l}$ properly.
To do this, we define the partition $\M_l=\M_{l,1} \sqcup \M_{l,2} \sqcup \ldots \sqcup \M_{l,k-1}$ where for all $j$:
$$\M_{l,j} = \left\{ i~\Big|~ p_i' \in \left[{1 \over k}+{(j-1)j \over 2}{1 \over k^2},{1 \over k}+{(j+1)j \over 2}{1 \over k^2} \right)\right\}.$$
(Notice that the length of interval used in the definition of $\M_{l,j}$ is $ {j \over k^2}$.) Now, for each $j=1,\ldots,k-1$ such that $\M_{l,j} \neq \emptyset$, we define $(q_i)_{i \in \M_{l,j}}$ via the following procedure:
\begin{enumerate}
\item Set $p_{j,\min}:={1 \over k}+{(j-1)j \over 2}{1 \over k^2}$, $p_{j,\max}:={1 \over k}+{(j+1)j \over 2}{1 \over k^2}$, $n_j = |\M_{l,j}|$, $\bar{p}_j = {\sum_{i \in {\cal M}_{l,j}}p_i' \over n_j}$.
\item Set $r=\left\lfloor \frac{n_j (\bar{p}_j-p_{j,\min})}{j/k^2}\right\rfloor$; let ${\cal M}_{l,j}' \subseteq {\cal M}_{l,j}$ be an arbitrary subset of cardinality $r$.
\item Set $q_i=p_{j,\max}$, for all $i \in {\cal M}_{l,j}'$; 
\item for an arbitrary index $i_j^* \in {\cal M}_{l,j} \setminus {\cal M}_{l,j}'$, set $q_{i^*_j}=n_j \bar{p}_j - (r p_{j,\max} + (n_j-r-1)p_{j,\min})$; 
\item finally, set $q_i=p_{j,\min}$, for all $i \in {\cal M}_{l,j} \setminus {\cal M}_{l,j}' \setminus \{i^*_j\}$.
\end{enumerate}
It is easy to see that
\begin{enumerate}
\item $\sum_{i \in {\cal M}_{l,j}}p_i' = \sum_{i \in {\cal M}_{l,j}}q_i \equiv n_j \bar{p}_j$;
\item for all $i \in {\cal M}_{l,j}\setminus \{i_j^*\}$, $q_i$ is an integer multiple of $1/k^2$.
\end{enumerate}
Moreover Theorem~\ref{thm: binomial approximation to the poisson binomial distribution} implies:
\begin{align*}\dtv{\sum_{i \in \M_{l,j}}Z_i}{\B\left(n_j,\bar{p}_j\right)} \le \frac{\sum_{i \in  \M_{l,j}} (p_i'-\bar{p}_j)^2}{(n_j+1) \bar{p}_j(1-\bar{p}_j)} &\le \begin{cases}{ n_j (j {1 \over k^2})^2    \over (n_j+1) p_{j,\min} (1-p_{j,\min})},~~~~\text{when $j <k-1$}\\{n_j (j {1 \over k^2})^2    \over (n_j+1) p_{j,\max} (1-p_{j,\max})},~~~~\text{when $j = k-1$}\end{cases}\\
&\le {8 \over k^2}.
\end{align*}
A similar derivation gives $\dtv{\sum_{i \in \M_{l,j}}Y_i}{\B\left(n_j,\bar{p}_j\right)} \le {8 \over k^2}$. So by the triangle inequality:
\begin{align}\dtv{\sum_{i \in \M_{l,j}}Z_i}{\sum_{i \in \M_{l,j}}Y_i} \le {16 \over k^2}. \label{eq: intermediate equation j}
\end{align}

\smallskip As~Eq~\eqref{eq: intermediate equation j} holds for all $j=1,\ldots,k-1$, an application of Lemma~\ref{lem:coupling} gives:
$$\dtv{\sum_{i \in \M_{l}}Z_i}{\sum_{i \in \M_{l}}Y_i} \le \sum_{j=1}^{k-1} \dtv{\sum_{i \in \M_{l,j}}Z_i}{\sum_{i \in \M_{l,j}}Y_i} \le {16 \over k}.$$
Moreover, the $q_i$'s defined above are integer multiples of $1/k^2$, except maybe for $q_{i_1^*},\ldots,q_{i_{k-1}^*}$. But we can round these to their closest multiple of $1/k^2$, increasing $\dtv{\sum_{i \in \M_l} Z_i}{\sum_{i \in \M_l}Y_i}$ by at most $1/k$.

\subsubsection{The Case $m > k^3$} \label{section: case m is large}
Let $t = | \{ i~|~p'_i=1\}|$. We show that the random variable $\sum_{i}Z_i$ is within total variation distance {$9/k$} from the Binomial distribution $\B(m',q)$ where
\begin{align*}
m':=\left\lceil \frac{\left(\sum_{i \in {\cal M}}p_i'  + t\right)^2}{\sum_{i \in {\cal M}}p_i'^2+t} \right\rceil \text{~~~and~~}&~~q:=\frac{\ell^*}{n},
\end{align*}
where $\ell^*$ satisfies $\frac{\sum_{i \in {\cal M}}p_i' + t}{m'} \in [\frac{\ell^*-1}{n}, \frac{\ell^*}{n}]$. Notice that:
\begin{itemize}
\item $\left(\sum_{i \in {\cal M}}p_i'  + t\right)^2 \le (\sum_{i \in {\cal M}}p_i'^2+t) (m+t)$, by the Cauchy-Schwarz inequality; and
\item $\frac{\sum_{i \in {\cal M}}p_i' + t}{m'} \le {\sum_{i \in \M} p_i' + t \over \frac{\left(\sum_{i \in {\cal M}}p_i'  + t\right)^2}{\sum_{i \in {\cal M}}p_i'^2+t}} = {\sum_{i \in {\cal M}}p_i'^2+t \over \sum_{i \in {\cal M}}p_i'+t} \le 1$.
\end{itemize}
So $m' \le m+t \le n$, and there exists some $\ell^* \in \{1,\ldots,n\}$ so that $\frac{\sum_{i \in {\cal M}}p_i' + t}{m'} \in [\frac{\ell^*-1}{n}, \frac{\ell^*}{n}]$.

\medskip For fixed $m'$ and $q$, we set $q_i=q$, for all $i \le m'$, and $q_i=0$, for all $i >m'$, and compare the distributions of  $\sum_{i \in {\cal M}}Z_i$ and $\sum_{i \in {\cal M}}Y_i$. For convenience we define 
\begin{align*}
\mu:=\E\left[\sum_{i \in {\cal M}}Z_i\right] &\text{~~and~~}\mu':=\E\left[\sum_{i \in {\cal M}}Y_i\right],\\
\sigma^2:=\text{Var}\left[\sum_{i \in {\cal M}}Z_i\right] &\text{~~and~~}\sigma'^2:=\text{Var}\left[\sum_{i \in {\cal M}}Y_i\right].
\end{align*}

\noindent The following lemma compares the values $\mu$, $\mu'$, $\sigma$, $\sigma'$.
\begin{lemma} \label{lemma: bounds on mus and sigma's}The following hold
\vspace{-8pt}
\begin{align}
\mu \le  \mu' \le \mu + 1,  \label{eq: mu difference}\\
\sigma^2-1 \le \sigma'^2 \le  \sigma^2 + 2, \label{eq: sigma square difference}\\
\mu \ge k^2, \label{eq: lower bound on mu}\\
\sigma^2 \ge  k^2\left(1-\frac{1}{k} \right).\label{eq: lower bound on sigma square}
\end{align}
\end{lemma}

\noindent The proof of Lemma~\ref{lemma: bounds on mus and sigma's} is given in Section~\ref{sec:missing proofs}. To compare $\sum_{i \in {\cal M}}Z_i$ and $\sum_{i \in {\cal M}}Y_i$ we approximate both by Translated Poisson distributions. Theorem~\ref{lem:translated Poisson approximation} implies that
\begin{align*}
\dtv{\sum_{i}Z_i}{TP(\mu, \sigma^2)} &\le \frac{\sqrt{\sum_{i}p_i'^3(1-p'_i)}+2}{\sum_{i}p'_i(1-p'_i)} \le \frac{\sqrt{\sum_{i}p_i'(1-p'_i)}+2}{\sum_{i}p'_i(1-p'_i)}\\
&\le \frac{1}{\sqrt{\sum_{i}p_i'(1-p'_i)}}+\frac{2}{\sum_{i}p'_i(1-p'_i)}= \frac{1}{\sigma}+\frac{2}{\sigma^2}\\
&\le \frac{1}{k\sqrt{1-1/k}}+\frac{2}{k^2\left(1-\frac{1}{k}\right)}~~~~~~~~\text{(using~\eqref{eq: lower bound on sigma square})}\\
& \le {3 \over k},
\end{align*}
where for the last inequality we assumed $k\ge 3$, but the bound of $3/k$ clearly also holds for $k=1,2$.
Similarly,
\begin{align*}
\dtv{\sum_{i}Y_i}{TP(\mu', \sigma'^2)} \le \frac{1}{\sigma'}+\frac{2}{\sigma'^2} &\le \frac{1}{k\sqrt{1-\frac{1}{k} - \frac{1}{k^2} }}+\frac{2}{k^2\left(1-\frac{1}{k} - \frac{1}{k^2} \right)}~~\text{(using~\eqref{eq: sigma square difference},\eqref{eq: lower bound on sigma square})}\\&\le {3 \over k},
\end{align*}
where for the last inequality we assumed $k\ge 3$, but the bound of $3/k$ clearly also holds for $k=1,2$. By the triangle inequality we then have that
\begin{align}
&\dtv{\sum_{i}Z_i}{\sum_{i }Y_i}\notag\\&~~~\le \dtv{\sum_{i }Z_i}{TP(\mu, \sigma^2)}+\dtv{\sum_{i}Y_i}{TP(\mu', \sigma'^2)} + \dtv{TP(\mu, \sigma^2)}{TP(\mu', \sigma'^2)} \notag\\
&~~~=6/k+\dtv{TP(\mu, \sigma^2)}{TP(\mu', \sigma'^2)}. \label{eq: tv triangle inequality stage 2a}
\end{align}
It remains to bound the total variation distance between the two Translated Poisson distributions. We make use of the following lemma.

\begin{lemma}[\cite{BarbourLindvall}] \label{lem: variation distance between translated Poisson distributions}
Let  $\mu_1, \mu_2 \in \mathbb{R}$ and $\sigma_1^2, \sigma_2^2 \in \mathbb{R}_+ \setminus \{0\}$ be such that $\lfloor \mu_1-\sigma_1^2 \rfloor \le \lfloor \mu_2-\sigma_2^2 \rfloor$. Then
\begin{align*}&\dtv{TP(\mu_1,\sigma_1^2)}{TP(\mu_2,\sigma_2^2)} \le \frac{|\mu_1-\mu_2|}{\sigma_1}+\frac{|\sigma_1^2-\sigma_2^2|+1}{\sigma_1^2}.\end{align*}
\end{lemma}
\noindent Lemma~\ref{lem: variation distance between translated Poisson distributions} implies
\begin{align}
&\dtv{TP(\mu, \sigma^2)}{TP(\mu', \sigma'^2)} \le \frac{|\mu-\mu'|}{\min(\sigma,\sigma')}+ \frac{|\sigma^2-\sigma'^2|+1}{\min(\sigma^2,\sigma'^2)} \notag\\
&~~~~~\le \frac{1}{k\sqrt{1-\frac{1}{k} - \frac{1}{k^2} }}+ \frac{3}{k^2\left(1-\frac{1}{k} - \frac{1}{k^2} \right)}~~~~\text{(using~Lemma~\ref{lemma: bounds on mus and sigma's})}\notag\\
&~~~~~\le 3/k,\label{eq:tv between the two translated Poisson's}
\end{align}
where for the last inequality we assumed $k > 3$, but the bound clearly also holds for $k=1,2,3$.
Using \eqref{eq: tv triangle inequality stage 2a} and \eqref{eq:tv between the two translated Poisson's} we get
\begin{align}
\dtv{\sum_{i}Z_i}{\sum_{i}Y_i} \le 9/k, \label{eq: stage 2(a) intermediate tv}
\end{align}
which implies~\eqref{eq: target equation Z,Y}.

\section{{Proof of Theorem~\ref{thm:moment matching}}}\label{sec: moment matching proof}

Let $\X$ and $\Y$ be two collections  of indicators as in the statement of Theorem~\ref{thm:moment matching}. For $\alpha_{\ell}(\cdot,\cdot)$ defined as in the statement of Theorem~\ref{thm:roos}, we claim the following.
\begin{lemma}\label{lemma:agreeing coefficients}
If $\P, \Q \in [0,1]^n$ satisfy property $(C_d)$ in the statement of Theorem~\ref{thm:moment matching}, then for all $p$, $\ell \in \{0,\ldots,d\}:$
$$\alpha_{\ell}(\mathcal{P},p) = \alpha_{\ell}(\mathcal{Q},p).$$
\end{lemma}
\begin{prevproof}{lemma}{lemma:agreeing coefficients}
First $\alpha_{0}(\mathcal{P},p) = 1 = \alpha_{0}(\mathcal{Q},p)$ by definition.
Now fix $\ell \in \{1,\ldots,d\}$ and consider the function $f(\vec{x}):=\alpha_{\ell}((x_1,\ldots,x_n),p)$ in the variables $x_1,\ldots,x_n \in \mathbb{R}$. Observe that $f$ is a symmetric polynomial of degree $\ell$ on $x_1,\ldots,x_n$. Hence, from the theory of symmetric polynomials, it follows that $f$ can be
written as a polynomial function of the power-sum symmetric polynomials
$\pi_1, \ldots ,\pi_{\ell}$, where 
$$\pi_j(x_1,\ldots,x_n):=\sum_{i=1}^nx_i^j,~\text{for all $j \in [\ell]$},$$
as the elementary symmetric polynomial of degree $j \in [n]$ can be written as a polynomial function of the power-sum symmetric polynomials $\pi_1,\ldots,\pi_j$ (e.g.~\cite{Zolotarev87}).
Now $(C_d)$ implies that $\pi_j(\P) = \pi_j(\Q)$, for all $j \le
\ell$. So $f(\P)=f(\Q)$, i.e.
$\alpha_{\ell}(\mathcal{P},p) = \alpha_{\ell}(\mathcal{Q},p)$.
\end{prevproof}

For all $p\in[0,1]$, by combining Theorem~\ref{thm:roos} and  Lemma~\ref{lemma:agreeing
coefficients} and  we get
that

\begin{align*}&Pr[X = m] - Pr[Y=m]= \sum_{\ell = d+1}^n (\alpha_{\ell}(\mathcal{P}, p)-\alpha_{\ell}(\mathcal{Q}, p))\cdot \delta^{\ell}\mathcal{B}_{n,p}(m),~\text{ for all $m \in \{0,\ldots,n\}$}.
\end{align*}
Hence, for all $p$:
\begin{align}
\dtv{X}{Y} &= \frac{1}{2} \sum_{m=0}^n|Pr[X = m] - Pr[Y=m]|\notag\\
&\le\frac{1}{2} \sum_{\ell = d+1}^n |\alpha_{\ell}(\mathcal{P}, p)-\alpha_{\ell}(\mathcal{Q}, p)|\cdot \| \delta^{\ell}\mathcal{B}_{n,p}(\cdot)\|_1\notag\\
&\le\frac{1}{2} \sum_{\ell = d+1}^n
\left(|\alpha_{\ell}(\mathcal{P}, p)|+|\alpha_{\ell}(\mathcal{Q},
p)|\right)\cdot \| \delta^{\ell}\mathcal{B}_{n,p}(\cdot)\|_1.
\label{eq:tv in terms of remaining terms coefficients}
\end{align}

Plugging $p = \bar{p}:= \frac{1}{n}\sum_{i} p_i$ into Proposition~\ref{prop:roos truncation quality}, we get
$$\theta(\P,\bar{p})= \frac{\sum_{i=1}^n(p_i - \bar{p})^2 }{n\bar{p}(1-\bar{p})} \le \Big|\max_i\{p_i\}-\min_i\{p_i\}\Big| \le \frac{1}{2}~~~~~~\text{(see~\cite{Roos00})}$$
and then
\begin{align*}
\frac{1}{2} \sum_{\ell = d+1}^n |\alpha_{\ell}(\mathcal{P},
\bar{p})|\cdot \| \delta^{\ell}\mathcal{B}_{n,\bar{p}}(\cdot)\|_1 &\le \sqrt{e}(d+1)^{1/4} 2^{-(d+1)/2} \frac{1-
\frac{1}{\sqrt{2}}\frac{d}{d+1}}{(\sqrt{2}-1)^2}\\& \le 6.5(d+1)^{1/4}
2^{-(d+1)/2}.
\end{align*}

But $(C_d)$ implies that $\sum_i q_i = \sum_i p_i  =\bar{p} $. So we get in a similar fashion
\begin{align*}
\frac{1}{2} \sum_{\ell = d+1}^n |\alpha_{\ell}(\mathcal{Q},
\bar{p})|\cdot \| \delta^{\ell}\mathcal{B}_{n,\bar{p}}(\cdot)\|_1
\le 6.5(d+1)^{1/4} 2^{-(d+1)/2}.
\end{align*}
Plugging these bounds into~\eqref{eq:tv in terms of remaining terms
coefficients} we get
$$\dtv{X}{Y} \le 13(d+1)^{1/4} 2^{-(d+1)/2}.$$

\section{{Deferred Proofs}} \label{sec:missing proofs}
\begin{prevproof}{Lemma}{lem: PBDs are fully determined by their probability vectors}
Let $X=\sum_i X_i$ and $Y=\sum_i Y_i$. It is obvious that, if $(p_1,\ldots,p_n)=(q_1,\ldots,q_n)$, then the distributions of $X$ and $Y$ are the same.  In the other direction, we show that, if $X$ and $Y$ have the same distribution, then $(p_1,\ldots,p_n)=(q_1,\ldots,q_n)$. 
%
%
%
Consider the polynomials:
\begin{align*}
g_X(s)=\E\left[ (1+s)^X\right] = \prod_{i=1}^n \E \left[ (1+s)^{X_i}\right] = \prod_{i=1}^n (1+p_i s);\\
g_Y(s)=\E\left[ (1+s)^Y\right] = \prod_{i=1}^n \E \left[ (1+s)^{Y_i}\right] = \prod_{i=1}^n (1+q_i s).
\end{align*}
Since $X$ and $Y$ have the same distribution,  $g_X$ and $g_Y$ are equal, so they have the same degree and roots. Notice that  $g_X$ has degree  $n-|\{i~|~p_i=0\}|$ and roots $\{ -{1 \over p_i}~|~p_i \neq 0 \}$. Similarly, $g_Y$ has degree $n-|\{i~|~q_i=0\}|$ and roots $\{ -{1 \over q_i}~|~q_i \neq 0 \}$. Hence, $(p_1,\ldots,p_n)=(q_1,\ldots,q_n)$.
\end{prevproof}

\begin{prevproof}{Lemma}{lem:coupling}
It follows from the coupling lemma that for any coupling of the variables $X_1,\ldots,X_n, Y_1,\ldots,Y_n$:
\begin{align}
\dtv{\sum_{i=1}^n X_i}{\sum_{i=1}^n Y_i} &\le \Pr\left[\sum_{i=1}^n X_i \neq \sum_{i=1}^n Y_i \right] \notag\\ &\le \sum_{i=1}^n \Pr[X_i \neq Y_i].\label{eq: coupling lemma}
\end{align}
We proceed to fix a specific coupling. For all $i$, it follows from the optimal coupling theorem that there exists a coupling of $X_i$ and $Y_i$ such that $\Pr[X_i \neq Y_i] = \dtv{X_i}{Y_i}$. Using these individual couplings for each $i$ we define a grand coupling of the variables $X_1,\ldots,X_n, Y_1,\ldots,Y_n$ such that $\Pr[X_i \neq Y_i] = \dtv{X_i}{Y_i}$, for all $i$. This coupling is faithful because $X_1,\ldots, X_n$ are mutually independent and $Y_1,\ldots,Y_n$ are mutually independent. Under this coupling~Eq~\eqref{eq: coupling lemma} implies:
\begin{align}
\dtv{\sum_{i=1}^n X_i}{\sum_{i=1}^n Y_i} \le \sum_{i=1}^n \Pr[X_i \neq Y_i] \equiv  \sum_{i=1}^n \dtv{X_i}{Y_i}.
\end{align}
\end{prevproof}

\begin{claim} \label{claim: easy to solve moment equations}
Fix integers $\tilde{n}, \delta, B, k \in \mathbb{N}_+$, $\tilde{n}, k\ge 2$. Given a set of values $\mu_1,\ldots,\mu_{\delta}, \mu'_1,\ldots,\mu'_{\delta}$, where, for all $\ell=1,\ldots,{\delta}$,
$$\mu_{\ell},\mu'_{\ell} \in \left\{0,\left({1 \over k^2}\right)^{\ell},2\left({1 \over k^2}\right)^{\ell},\ldots,B\right\},$$
discrete sets $\mathcal{T}_1,\ldots,\mathcal{T}_{\tilde{n}} \subseteq \left\{0, {1 \over k^2}, {2 \over k^2},\ldots, 1\right\}$, and four integers $n_0, n_1 \le {\tilde{n}}$, $n_s, n_b \le B$, it is possible to solve the system of equations:
\begin{align*}
(\Sigma):~~~\sum_{p_i \in (0,1/2]}p_i^{\ell}&=\mu_{\ell}, \text{ for all $\ell=1,\ldots,{\delta}$},\\
\sum_{p_i \in (1/2,1)}p_i^{\ell}&=\mu'_{\ell}, \text{ for all $\ell=1,\ldots,{\delta}$},\\
|\{i | p_i = 0\}|&= n_0\\
|\{i | p_i = 1\}|&=n_1\\
|\{i | p_i \in (0, 1/2]\}|&=n_s\\
|\{i | p_i \in (1/2, 1)\}|&=n_b
\end{align*}
 with respect to the variables $p_1 \in \mathcal{T}_1,\ldots, p_{\tilde{n}} \in \mathcal{T}_{\tilde{n}}$, or to determine that no solution exists, in time
$$O({\tilde{n}}^3 \log_2 {\tilde{n}}) B^{O({\delta})} k^{O({\delta}^2)}.$$
\end{claim}

\begin{prevproof}{Claim}{claim: easy to solve moment equations}
We use dynamic programming. Let us consider the following tensor of
dimension $2{\delta}+5$:
$$A(i, z_0, z_1, z_s, z_b; \nu_1,\ldots,\nu_{\delta} ; \nu'_1,\ldots,\nu'_{\delta}),$$
where $i \in [\tilde{n}]$, $z_0, z_1 \in \{0,\ldots,\tilde{n}\}$, $z_s, z_b \in \{0,\ldots,B\}$ and $$\nu_{\ell},
\nu'_{\ell} \in \left\{0,\left({1 \over k^2}\right)^{\ell},2\left({1
\over k^2}\right)^{\ell},\ldots,B\right\},$$ for $\ell=1,\ldots,{\delta}.$
The total number of cells in $A$ is
\begin{align*}&\tilde{n} \cdot (\tilde{n}+1)^2 \cdot (B+1)^2 \cdot \left( \prod_{\ell =1}^{\delta} (B k^{2 \ell}+1) \right)^2 \le O(\tilde{n}^3) B^{O({\delta})} k^{O({\delta}^2)}.
\end{align*}
Every cell of $A$ is assigned value $0$ or $1$, as follows:
\begin{align*}
&A(i, z_0, z_1, z_s, z_b ; \nu_1,\ldots,\nu_{\delta}, \nu'_1,\ldots,\nu'_{\delta})=1\\&\\&~~\Longleftrightarrow \left(\begin{minipage}{6.5cm} \centering There exist $p_1 \in \mathcal{T}_1$, $\ldots$, $p_i \in \mathcal{T}_i$ such that $|\{j \le i | p_j =0\}|=z_0$, $|\{j \le i | p_j =1\}|=z_1,$ $|\{j \le i | p_j \in (0,1/2]\}|=z_s$, $|\{j \le i | p_j \in (1/2,1)\}|=z_b,$
$\sum_{j \le i: p_j \in (0,1/2]}p_j^{\ell}=\nu_{\ell}$, for all
$\ell=1,\ldots,{\delta}$, $\sum_{j \le i: p_j \in
(1/2,1)}p_j^{\ell}=\nu'_{\ell}$, for all
$\ell=1,\ldots,{\delta}$.\end{minipage} \right).
\end{align*} 

Notice that we need $O(\tilde{n}^3) B^{O({\delta})} k^{O({\delta}^2)}$ bits to store $A$ and $O(\log \tilde{n}+ \delta \log B + \delta^2 \log k)$ bits to address cells of $A$. To complete
$A$ we can work in layers of increasing $i$. We initialize all entries
to value $0$. Then, the first layer
$A(1,\cdot,\cdot~;~\cdot,\ldots,\cdot)$ can be completed easily as
follows:
\begin{align*}
&A(1, 1, 0, 0, 0 ; 0, 0, \ldots,0 ; 0, 0, \ldots, 0)=1 \Leftrightarrow 0 \in \mathcal{T}_1;\\
&A(1, 0, 1, 0, 0 ; 0, 0, \ldots,0 ; 0, 0 \ldots, 0)=1 \Leftrightarrow 1 \in \mathcal{T}_1;\\
&A(1, 0, 0, 1, 0 ; p, p^2, \ldots,p^{\delta} ; 0,\ldots,0)=1 \Leftrightarrow p \in \mathcal{T}_1 \cap (0,1/2];\\
&A(1, 0, 0, 0, 1 ; 0, \ldots,0; p, p^2, \ldots,p^{\delta})=1 \Leftrightarrow p \in \mathcal{T}_1 \cap (1/2,1).
\end{align*}
Inductively, to complete layer $i+1$, we consider all the non-zero
entries of layer $i$ and for every such non-zero entry and for every
$v_{i+1} \in \mathcal{T}_{i+1}$, we find which entry of layer $i+1$
we would transition to if we chose $p_{i+1}=v_{i+1}$. We set that
entry equal to $1$ and we also save a pointer to this entry from the
corresponding entry of layer $i$, labeling that pointer with the
value $v_{i+1}$. The bit operations required to complete layer $i+1$ are bounded
by
$$|\mathcal{T}_{i+1}| (\tilde{n}+1)^2 B^{O({\delta})} k^{O({\delta}^2)} O(\log \tilde{n} + \delta \log B + \delta^2 \log k) \le O(\tilde{n}^2 \log \tilde{n}) B^{O({\delta})} k^{O({\delta}^2)}.$$
Therefore, the overall time needed to complete $A$ is $$O(\tilde{n}^3 \log \tilde{n})
B^{O({\delta})} k^{O({\delta}^2)}.$$

Having completed $A$, it is easy to check if there is a
solution to $(\Sigma)$. A solution exists if and only if
$$A(\tilde{n},{n}_0,n_1,n_s,n_b;\mu_1,\ldots,\mu_{\delta} ; \mu'_1,\ldots,\mu'_{\delta})=1,$$ and can
be found by tracing the pointers from this cell of $A$ back to level $1$. The
overall running time is dominated by the time needed to complete $A$.
\end{prevproof}

\begin{prevproof}{lemma}{lem:variation distance between Poisson distributions}
Without loss of generality assume that $0<\lambda_1 \le \lambda_2$ and denote $\delta = \lambda_2-\lambda_1$. For all $i\in\{0,1,\ldots\}$, denote 
$$p_i=e^{-\lambda_1}\frac{\lambda_1^i}{i!} \text{ ~~~~ and~~~   }q_i=e^{-\lambda_2}\frac{\lambda_2^i}{i!}.$$
Finally, define $\I^*=\{i:p_i\ge q_i\}$.\\

\noindent We have
\begin{align*}
\sum_{i \in \I^*}{|p_i - q_i|} = \sum_{i \in \I^*}{(p_i - q_i)}  &\le \sum_{i \in \I^*}{\frac{1}{i!}(e^{-\lambda_1}\lambda_1^i - e^{-\lambda_1-\delta}\lambda_1^i)}\\
&= \sum_{i \in \I^*}{\frac{1}{i!}e^{-\lambda_1}\lambda_1^i(1 - e^{-\delta})}\\
&\le (1 - e^{-\delta}) \sum_{i=0}^{+\infty}{\frac{1}{i!}e^{-\lambda_1}\lambda_1^i}=1 - e^{-\delta}.
\end{align*}

\noindent On the other hand
\begin{align*}
\sum_{i \notin \I^*}{|p_i - q_i|} = \sum_{i \notin \I^*}{(q_i - p_i)}  &\le \sum_{i \notin \I^*}{\frac{1}{i!}(e^{-\lambda_1}(\lambda_1+\delta)^i - e^{-\lambda_1}\lambda_1^i)}\\
&= \sum_{i \notin \I^*}{\frac{1}{i!}e^{-\lambda_1}((\lambda_1+\delta)^i-\lambda_1^i)}\\
&\le \sum_{i=0}^{+\infty}{\frac{1}{i!}e^{-\lambda_1}((\lambda_1+\delta)^i-\lambda_1^i)}\\
&= e^{\delta}\sum_{i=0}^{+\infty}{\frac{1}{i!}e^{-(\lambda_1+\delta)}(\lambda_1+\delta)^i }- \sum_{i=0}^{+\infty}{\frac{1}{i!}e^{-\lambda_1}\lambda_1^i}\\
&= e^{\delta}-1.
\end{align*}
Combining the above we get the result.
\end{prevproof}

\begin{prevproof}{lemma}{lemma: bounds on mus and sigma's}
We have
$${\mu \over m'}=\frac{\sum_{i \in {\cal M}}p_i'+t}{m'} \le q=\frac{\ell^*}{n} \le \frac{\sum_{i \in {\cal M}}p_i'+t}{m'}+\frac{1}{n} = {\mu \over m'}+{1 \over n}.$$
Multiplying by $m'$ we get:
$${\mu} \le m' q \le \mu +{m' \over n}.$$
As $\mu' = m' q$ and $m' \le n$, we get ${\mu} \le \mu' \le \mu + 1.$ Moreover, since $m\ge k^3$, 
$$\mu \ge \sum_{i \in {\cal M}}p'_i \ge m \frac{1}{k} \ge k^2.$$

For the variances we have:
\begin{align}
\sigma'^2 = m' q (1-q) &= m' \cdot {\ell^* \over n} \cdot \left(1- {\ell^*-1 \over n} -{1 \over n}\right)\notag\\
&\ge \left(\sum_{i \in \M}p_i' +t \right) \cdot \left(1- {1 \over n} - {\sum_{i \in \M} p_i' +t \over m'} \right)\notag\\
&= (1-1/n) \left(\sum_{i \in \M}p_i' +t \right)  - {(\sum_{i \in \M} p_i' +t)^2 \over m'}\notag\\
&\ge (1-1/n) \left(\sum_{i \in \M}p_i' +t \right)  - {(\sum_{i \in \M} p_i' +t)^2 \over \frac{\left(\sum_{i \in {\cal M}}p_i'  + t\right)^2}{\sum_{i \in {\cal M}}p_i'^2+t}}\notag\\
&=  \sum_{i \in {\cal M}}p'_i(1-p'_i)  -{1 \over n} \left(\sum_{i \in \M}p_i' +t \right)\notag\\
&= \sigma^2  -{1 \over n} \left(\sum_{i \in \M}p_i' +t \right) \ge \sigma^2  -1.
\end{align}
In the other direction:
\begin{align}
\sigma'^2 = m' q (1-q) &= m' \cdot \left({\ell^* -1 \over n} + {1 \over n} \right) \cdot \left(1- {\ell^* \over n} \right) \notag\\
&\le m' \cdot \left({\ell^* -1 \over n}\right) \cdot \left(1- {\ell^* \over n} \right)+{m' \over n} \notag\\
&\le \left({\sum_{i \in {\cal M}}p_i' + t}\right)\cdot \left(1- \frac{\sum_{i \in {\cal M}}p_i' + t}{m'}\right)+1\notag\\
&= \left({\sum_{i \in {\cal M}}p_i' + t}\right) - \frac{\left(\sum_{i \in {\cal M}}p_i' + t \right)^2}{m'}+1\notag\\
&\le \left({\sum_{i \in {\cal M}}p_i' + t}\right) - \frac{\left(\sum_{i \in {\cal M}}p_i' + t \right)^2}{{{\left(\sum_{i \in {\cal M}}p_i'  + t\right)^2} \over {\sum_{i \in {\cal M}}p_i'^2+t}} +1 }+1\notag\\
&= \left({\sum_{i \in {\cal M}}p_i' + t}\right) - \left({\sum_{i \in {\cal M}}p_i'^2+t}\right)\frac{\left(\sum_{i \in {\cal M}}p_i' + t \right)^2}{{{\left(\sum_{i \in {\cal M}}p_i'  + t\right)^2}} +{\sum_{i \in {\cal M}}p_i'^2+t} }+1\notag\\
&= \left({\sum_{i \in {\cal M}}p_i' + t}\right) - \left({\sum_{i \in {\cal M}}p_i'^2+t}\right)\left(1-{\sum_{i \in {\cal M}}p_i'^2 + t  \over {{{\left(\sum_{i \in {\cal M}}p_i'  + t\right)^2}} +{\sum_{i \in {\cal M}}p_i'^2+t} }}\right)+1\notag\\
&= {\sum_{i \in {\cal M}}p_i' (1-p_i')} + {\left(\sum_{i \in {\cal M}}p_i'^2 + t \right)^2 \over {{{\left(\sum_{i \in {\cal M}}p_i'  + t\right)^2}} +{\sum_{i \in {\cal M}}p_i'^2+t} }}+1\notag\\
&= \sigma^2 + {\left(\sum_{i \in {\cal M}}p_i'^2 + t \right)^2 \over {{{\left(\sum_{i \in {\cal M}}p_i'  + t\right)^2}} +{\sum_{i \in {\cal M}}p_i'^2+t} }}+1 \le \sigma^2+2.
\end{align}
Finally,
$$\sigma^2 = \sum_{i \in {\cal M}}p'_i(1-p'_i) \ge m \frac{1}{k}\left(1-\frac{1}{k} \right)\ge k^2\left(1-\frac{1}{k} \right).$$
\end{prevproof}

\begin{prop}\label{proposition:variable moments to probability moments}
For all $d \in [n]$, Condition $(C_d)$ in the statement of Theorem~\ref{thm:moment matching} is equivalent to the following condition:

$$(V_d):~~\E\left[\left(\sum_{i=1}^n X_i\right)^{\ell}\right] = \E\left[\left(\sum_{i=1}^n Y_i\right)^{\ell}\right],~\text{for all } \ell \in [d].$$
\end{prop}
\begin{prevproof}{Proposition}{proposition:variable moments to probability moments}\text{}\\
$(V_d) \Rightarrow (C_d)$:~First notice that, for all $\ell \in [n]$,
$\E\left[\left(\sum_{i=1}^n X_i\right)^{\ell}\right]$ can be written
as a weighted sum of the {elementary symmetric polynomials}
$\psi_1(\P)$, $\psi_2(\P)$,...,$\psi_{\ell}(\P)$, 
where, for all $t \in [n]$, $\psi_{t}(\P)$ is defined as
$$\psi_{t}(\P):=(-1)^{t}\sum_{\begin{minipage}{1.5cm}\centering$S \subseteq [n]$\\$|S|=t$\end{minipage}} \prod_{i \in S} p_i.$$
$(V_d)$  implies then by induction
\begin{align}\psi_{\ell}(\P)=\psi_{\ell}(\Q),~~~\text{for all $\ell = 1,\ldots,d$}.\label{eq: equality of the elementary symmetric polynomials}
\end{align}
Next, for all $t \in [n]$, define $\pi_{t}(\P)$ to be the power sum
symmetric polynomial of degree $t$
$$\pi_{t}(\mathcal{P}):=\sum_{i=1}^np_i^{t}.$$
Now, fix any $\ell \le d$. Since $\pi_{\ell}(\P)$ is a
symmetric polynomial of degree $\ell$ on the variables
$p_1,\ldots,p_n$, it  can be expressed as a function of the
elementary symmetric polynomials
$\psi_1(\P),\ldots,\psi_{\ell}(\P)$. So, by~\eqref{eq: equality
of the elementary symmetric polynomials}, $\pi_{\ell}(\mathcal{P}) =
\pi_{\ell}(\mathcal{Q})$. Since this holds for any $\ell \le d$,
$(C_d)$ is satisfied.

\medskip \noindent The implication $(C_d) \Rightarrow (V_d)$ is established in a
similar fashion. $(C_d)$ says that
\begin{align}
\pi_{\ell}(\mathcal{P})=\pi_{\ell}(\mathcal{Q}), \text{for all $\ell=1,\ldots,d$}.\label{eq: power sum are equal}
\end{align}
Fix some $\ell \le d$. $\E\left[\left(\sum_{i=1}^n
X_i\right)^{\ell}\right]$ can be written
as a weighted sum of the elementary symmetric polynomials
$\psi_1(\P)$, $\psi_2(\P)$,...,$\psi_{\ell}(\P)$. Also, for all $t \in [\ell]$, $\psi_t(\P)$ can be written as a polynomial function of $\pi_1(\P),\ldots,\pi_{t}(\P)$ (see, e.g.,~\cite{Zolotarev87}). So from~\eqref{eq: power sum are equal} it follows that $\E\left[\left(\sum_{i=1}^n
X_i\right)^{\ell}\right]=\E\left[\left(\sum_{i=1}^n
Y_i\right)^{\ell}\right]$. Since this holds for any $\ell \le d$,
$(V_d)$ is satisfied.
\end{prevproof}

\begin{corollary} \label{theorem:binomial appx theorem for large guys}
Let $\mathcal{P}:=(p_i )_{i=1}^n \in [1/2,1]^n$ and $\mathcal{Q}:=(q_i)_{i=1}^n \in [1/2,1]^n$ be two collections of  probability values in $[1/2,1]$. Let also $\mathcal{X}:=(X_i)_{i=1}^n$ and $\mathcal{Y}:=(Y_i)_{i=1}^n$ be two collections of mutually independent indicators with $\E[X_i]=p_i$ and $\E[Y_i]=q_i$, for all $i \in [n]$. If for some $d \in [n]$ Condition $(C_d)$ in the statement of Theorem~\ref{thm:moment matching} is satisfied, then
$$\dtv{\sum_{i}{X_i}}{\sum_{i}{Y_i}} \le 13(d+1)^{1/4} 2^{-(d+1)/2}.$$
\end{corollary}

\begin{prevproof}{Corollary}{theorem:binomial appx theorem for large guys}
Define $X'_i = 1-X_i$ and $Y'_i=1-Y_i$, for all $i$. Also, denote $p_i'=\mathbb{E}[X'_i]=1-p_i$ and $q_i'=\mathbb{E}[Y'_i]=1-q_i$, for all $i$. By assumption:
\begin{align}
\sum_{i=1}^n \left(1-p_i'\right)^{\ell} = \sum_{i=1}^n \left(1-q_i'\right)^{\ell},~~~\text{for all } \ell=1,\ldots,d. \label{eq:hypothesis corollary}
\end{align}
Using the Binomial theorem and induction, we see that \eqref{eq:hypothesis corollary} implies:
$$\sum_{i=1}^n p_i'^{\ell} = \sum_{i=1}^n q_i'^{\ell},~~~\text{for all } \ell=1,\ldots,d.$$
Hence we can apply
Theorem~\ref{thm:moment matching} to deduce
$$\dtv{\sum_{i}{X'_i}}{\sum_{i}{Y'_i}}\le 13(d+1)^{1/4} 2^{-(d+1)/2}.$$
The proof is completed by noticing that
$$\dtv{\sum_{i}{X_i}}{\sum_{i}{Y_i}}=\dtv{\sum_{i}{X'_i}}{\sum_{i}{Y'_i}}.$$
\end{prevproof}

\section*{Acknowledgement}
We thank the anonymous reviewer for comments that helped improve the presentation. 
\bibliographystyle{alpha}
\bibliography{costasbib}

\end{document}